\documentclass[prx,preprint,floatfix,superscriptaddress,amsmath,aps,nolongbibliography]{revtex4-2}
\usepackage{dcolumn,color,booktabs,microtype,afterpage,bm}

\usepackage[colorlinks,plainpages=false,linkcolor=blue,urlcolor=blue,citecolor=blue,pdfpagemode=UseNone,pdfstartview=FitBH]{hyperref}
\usepackage{amsmath}
\usepackage{bm}
\usepackage{tabularx}
\usepackage{graphicx}
\usepackage{lineno}
\usepackage{enumerate}
\usepackage{multirow}
\usepackage{color}
\usepackage{mhchem}
\usepackage{comment}

\begin{document}
\newcommand{\eiig}{EuIr$_4$In$_2$Ge$_4$}
\newcommand{\tn}{$T_{\rm N}$~}
\newcommand{\mub}{${\mu}_{B}$~}
\newcommand{\QQ}{${\bm Q}$}
\newcommand{\qvec}{${\bm q}$}

\sloppy
\title{Magnetic Order of Dresselhaus-type Antiferromagnet \eiig~Studied by Single Crystal Neutron Diffraction}

\author{Chihiro Tabata$^\dagger$}
\affiliation{Materials Sciences Research Center, Japan Atomic Energy Agency, Tokai 319-1195, Japan}
\affiliation{Advanced Science Research Center, Japan Atomic Energy Agency, Tokai 319-1195, Japan}

\author{Koji Kaneko}
\affiliation{Materials Sciences Research Center, Japan Atomic Energy Agency, Tokai 319-1195, Japan}
\affiliation{Advanced Science Research Center, Japan Atomic Energy Agency, Tokai 319-1195, Japan}
\affiliation{J-PARC center, Japan Atomic Energy Agency, Tokai 319-1195, Japan}

\author{Akiko Nakao}
\affiliation{Comprehensive Research Organization for Science and Society, Tokai, Ibaraki 319-1106, Japan}

\author{Takashi Ohhara}
\affiliation{J-PARC center, Japan Atomic Energy Agency, Tokai 319-1195, Japan}

\author{Tatsuma D. Matsuda}
\affiliation{Department of Physics, Tokyo Metropolitan University, Hachioji, Tokyo 192-0397, Japan}

\author{Yoshichika \={O}nuki}
\affiliation{Department of Physics, Tokyo Metropolitan University, Hachioji, Tokyo 192-0397, Japan}
\affiliation{RIKEN Center for Emergent Matter Science, Wako, Saitama 352-0198, Japan}


\begin{abstract}
The magnetic order of \eiig, which crystallizes in a Dresselhaus-type noncentrosymmetric tetragonal structure, was investigated using two complementary single-crystal neutron diffraction approaches. 
Time-of-flight single-crystal diffraction reveals antiferromagnetic Bragg reflections with propagation vector $\bm{q} = (1, 0, 0)$ below the N\'{e}el temperature \tn = 2.5~K, indicating a breaking of body-centered translational symmetry.
Polarized neutron diffraction on a triple-axis spectrometer demonstrates that the ordered \ce{Eu^2+} $4f$ moments lie within the basal plane and form a collinear antiferromagnetic structure with antiparallel alignment between corner and body-center sites.
Despite the Dresselhaus-type spin splitting in the conduction bands, the magnetic order remains simple, implying weak coupling between localized moments and itinerant electrons.

\end{abstract}

\maketitle

In noncentrosymmetric systems, nontrivial properties such as topological magnetic orders and anomalous transport phenomena can emerge through antisymmetric spin-orbit coupling (SOC). Representative examples include Rashba-type SOC, which arises from broken mirror symmetry, and Dresselhaus-type SOC, which originates from inversion-symmetry breaking in bulk crystals \cite{Dresselhaus1955-bg}.
The former has been widely studied not only in numerous heterostructures but also in bulk compounds such as BiAs \cite{Zubair2024-qr}, \ce{CePt3Si} \cite{Bauer2004-nz} and \ce{CeIrSi3} \cite{Onuki2012-om}, whereas only a limited number of examples of the latter have been reported.

The intermetallic semiconductor {\eiig} is one of the few known materials that exhibit a Dresselhaus-type spin splitting.
This compound crystallizes in a noncentrosymmetric tetragonal structure with space group $I\bar{4}2m$ (No. 121, $D_{2d}^{11}$). 
Band-structure calculations indicate that the band gap originates predominantly from Ir 5$d$ electrons, with small contributions from Ge 4$p$ and In 5$p$ electrons, and that the combination of strong SOC of Ir and the broken inversion symmetry leads to a pronounced Dresselhaus-type spin splitting \cite{Calta2015-tw, Calta2016-cn}. 
This characteristic electronic structure gives rise to unconventional magnetic transport phenomena.
In particular, recent experiments using microscale single-crystal samples fabricated by focused ion beam have revealed nonreciprocal magnetoresistance with a magnetic-field-angle dependence consistent with the symmetry of the Dresselhaus-type band splitting \cite{Yokoyama2025-aa}.

In contrast to the characteristic band structure, the magnetism of {\eiig} has remained largely unexplored. 
Bulk property measurements indicate that divalent Eu ions carry localized magnetic moments and undergo antiferromagnetic ordering at \tn = 2.5 K \cite{Nakachi2024-yg}, whereas microscopic information has so far been lacking.
Band-structure calculations predict that the \ce{Eu^2+} moments do not contribute to the spin splitting of the conduction band \cite{Calta2015-tw,Calta2016-cn}; nevertheless, the nonreciprocal magnetoresistance is already strongly suppressed above \tn \cite{Yokoyama2025-aa}.
In addition, thermal conductivity measurements have revealed a pronounced anisotropy that is absent in the nonmagnetic analogue \ce{SrIr4In2Ge4} \cite{Stachowiak2023-kx}. 
These observations suggest a coupling between conduction electrons exhibiting Dresselhaus-type spin splitting and localized magnetic moments, making the resulting magnetic order of particular interest. 

In this study, we investigated the magnetic order of {\eiig} by single-crystal neutron diffraction.
Despite the presence of highly neutron-absorbing elements, such as Eu, Ir and In, the magnetic structure was successfully revealed by employing complementary single-crystal neutron diffraction techniques.
The antiferromagnetic order of {\eiig} is characterized by a propagation vector $\bm{q} = (1, 0, 0)$, with the \ce{Eu^2+} magnetic moments confined to the basal plane. 
In contrast to the Dresselhaus-type spin textures of the conduction electrons, the \ce{Eu^2+} moments form a simple collinear magnetic structure.

Single crystalline {\eiig} was grown by the In-self flux method, the details of which have been reported in Ref. \onlinecite{Nakachi2024-yg}. 
A cube-like single crystal with typical dimensions of 2 $\times$ 2 $\times$ 2 mm$^3$ was used for the neutron diffraction experiments described below.

\begin{figure*}[t]
\centering
        \includegraphics[width=18cm]{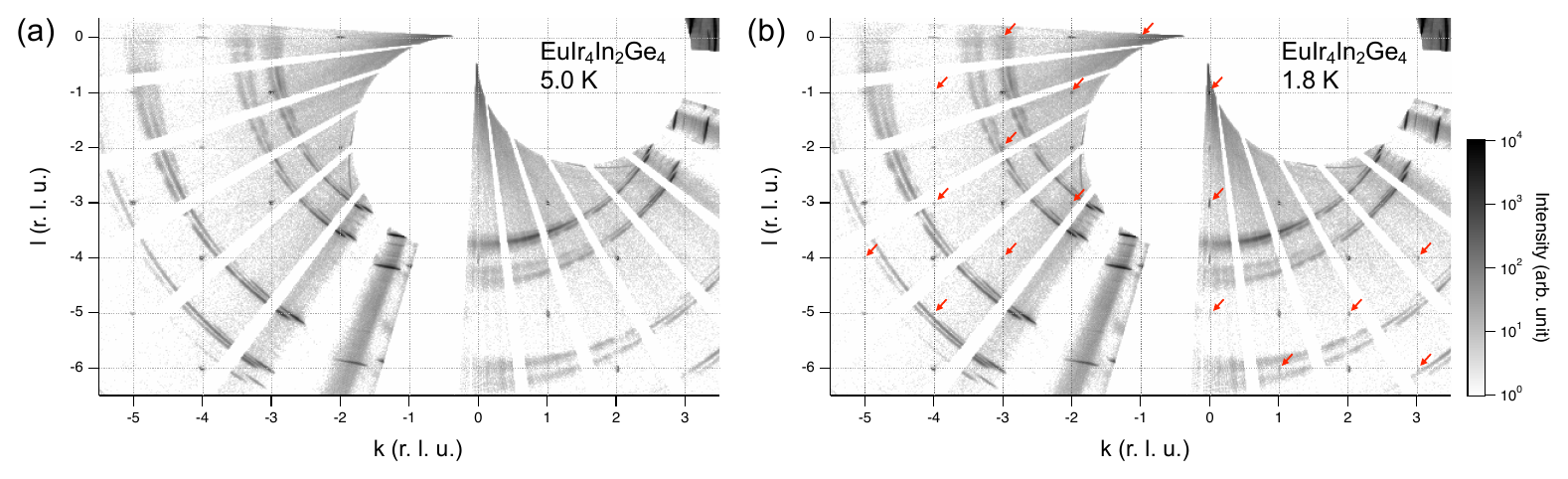}
    \caption{(Color online) Neutron scattering intensity maps of the reciprocal (0, $k$, $l$) plane recorded at two sample orientations at (a)~5.0~K ($T > T_{\rm N}$) and (b)~1.8~K ($T <T_{\rm N}$). The arrows in (b) indicate the additional peaks observed below {\tn}. }
     \label{fig:map}
\end{figure*}

Single-crystal neutron diffraction measurements were first carried out using the time-of-flight Laue diffractometer SENJU installed at BL18 at the Materials and Life Science Facility (MLF), J-PARC, Tokai \cite{Ohhara2016-ge}. 
The first frame, covering a neutron wavelength range from 0.4 to 4.0~{\AA}, was employed to access a wide region of reciprocal space.
The single-crystal sample was glued to an aluminum plate, mounted on a sample stick, and loaded into a top-load cryostat.
Data were collected at 1.8~K ($T < T_{\rm N}$) and at 5.0~K ($T > T_{\rm N}$).
Data reduction and reciprocal-space mapping were performed using the software package STARGazer \cite{Ohhara2016-ge}.

After identifying the magnetic propagation vector at SENJU, further investigations were performed on the thermal triple-axis spectrometer TAS-1 at the JRR-3 research reactor.
The spectrometer was operated in the polarized triple-axis mode with a fixed wavelength of 2.36~{\AA}.
The same single-crystal sample used at SENJU was mounted on TAS-1 with ($h,k,0$) plane as the horizontal scattering plane.
The sample was cooled down to 0.28~K using a $^3$He insert installed in a variable-temperature space of a dry top-load-type cryostat \cite{Kaneko2024-aq}.
A double-focusing Heusler monochromator and analyzer were used with a collimation sequence of open-80'-80'-80', and higher-order contamination was removed by a pyrolytic graphite (PG) filter placed before the sample.

Neutron polarization was controlled using a Helmholtz coil together with the guide field, allowing polarization parallel or antiparallel to the scattering vector {\QQ} ($P_x$ mode), and perpendicular to both {\QQ} and horizontal scattering plane ($P_z$ mode).
Using a Mezei-type spin flipper installed upstream of the sample, the flipping ratio (FR) was evaluated from the 2~0~0 nuclear Bragg reflection, as described in the following section.
Superlattice reflections associated with magnetic order were subsequently investigated.

Superlattice reflections associated with magnetic order were surveyed over a wide region of reciprocal space using the time-of-flight Laue method on SENJU.
Figure \ref{fig:map} shows representative neutron diffraction intensity maps in the ($0,k,l$) scattering plane recorded at (a) 5.0~K ($T > T_{\rm N}$) and (b) 1.8~K ($T <T_{\rm N}$).
In the paramagnetic state at 5.0~K, nuclear Bragg reflections were observed at $k+l={\rm even}$, consistent with the extinction rules of the space group $I{\bar 4}2m$.

Upon cooling below {\tn} to 1.8~K, additional reflections appeared at integer positions in the ($0,k,l$) plane with $k+l={\rm odd}$ as indicated by red arrows in Fig.~\ref{fig:map}(b).
These observations indicate that the propagation vector is $\bm{q}=(1, 0, 0)$, implying that the magnetic order breaks the body-centered translational symmetry of the crystal lattice.
A notable feature is the presence of intense magnetic reflections along the series of $0~0~l$ with $l=$ odd, such as 0~0~1, 0~0~3, and 0~0~5. 
Since magnetic scattering is sensitive only to magnetic-moment components perpendicular to the scattering vector {\QQ}, this result provides definite evidence that the ordered magnetic moments have an in-plane component.

Since the propagation vector was determined as $\bm{q}=(1, 0, 0)$, further insight into the magnetic order was obtained using polarized neutron scattering on a triple-axis spectrometer.
A conventional magnetic structure refinement based solely on integrated intensities is highly challenging for {\eiig} because of its strong neutron absorption; Eu, Ir and In possess large absorption cross sections for thermal neutrons, ${\sigma}_a =$ 4,530, 425, and 194 barns, respectively \cite{Varley_F1992-je}.
In this context, polarized neutron scattering provides direct and reliable information on the orientation of the ordered moments without requiring absorption corrections, which is particularly advantageous for examining possible canting of the moments away from the basal plane.

For the present longitudinal polarization analysis, the neutron scattering intensities in the $P_x$ and $P_z$ modes are expressed as

    \begin{gather}
        P_x^{\rm NSF}{\propto}~N  \\
        P_x^{\rm SF}{\propto}~M_{\perp}+M_{\parallel} {\pm} M_{\rm ch}  \\
        P_z^{\rm NSF}{\propto}~M_{\parallel} \\
        P_z^{\rm SF}{\propto}~M_{\perp} 
    \end{gather}

where $N$ denotes the nuclear scattering intensity, and $M_{\perp}$ and $M_{\parallel}$ represent the magnetic scattering intensities arising from moment components perpendicular and parallel to the $c$-axis, respectively.
The term $M_{\rm ch}$ is a chiral term that becomes nonzero only for noncollinear spin arrangements with a finite chirality.

Strictly speaking, these expressions are valid only for an ideal instrument with perfect polarization efficiency between SF and NSF channels.
In practice, each polarization channel inevitably contains a finite contamination from the other.
Therefore, the measured intensities were corrected using the instrumental flipping ratio $R$, determined from a reference nuclear Bragg reflection.
The corrected SF and NSF intensities, $I_{\rm SF}^{\rm cor}$ and $I_{\rm NSF}^{\rm cor}$, are given by
\begin{gather}
    I_{\rm SF}^{\rm cor} = \frac{1}{R-1}(RI_{\rm SF}^{\rm obs}-I_{\rm NSF}^{\rm obs} ) \\
    I_{\rm NSF}^{\rm cor} = \frac{1}{R-1}(RI_{\rm NSF}^{\rm obs}-I_{\rm SF}^{\rm obs})
\end{gather}
where $I_{\rm SF}^{\rm obs}$ and $I_{\rm NSF}^{\rm obs}$ denote the uncorrected SF and NSF intensities, respectively.

\begin{figure}[b]
\centering
        \includegraphics[width=7.5cm]{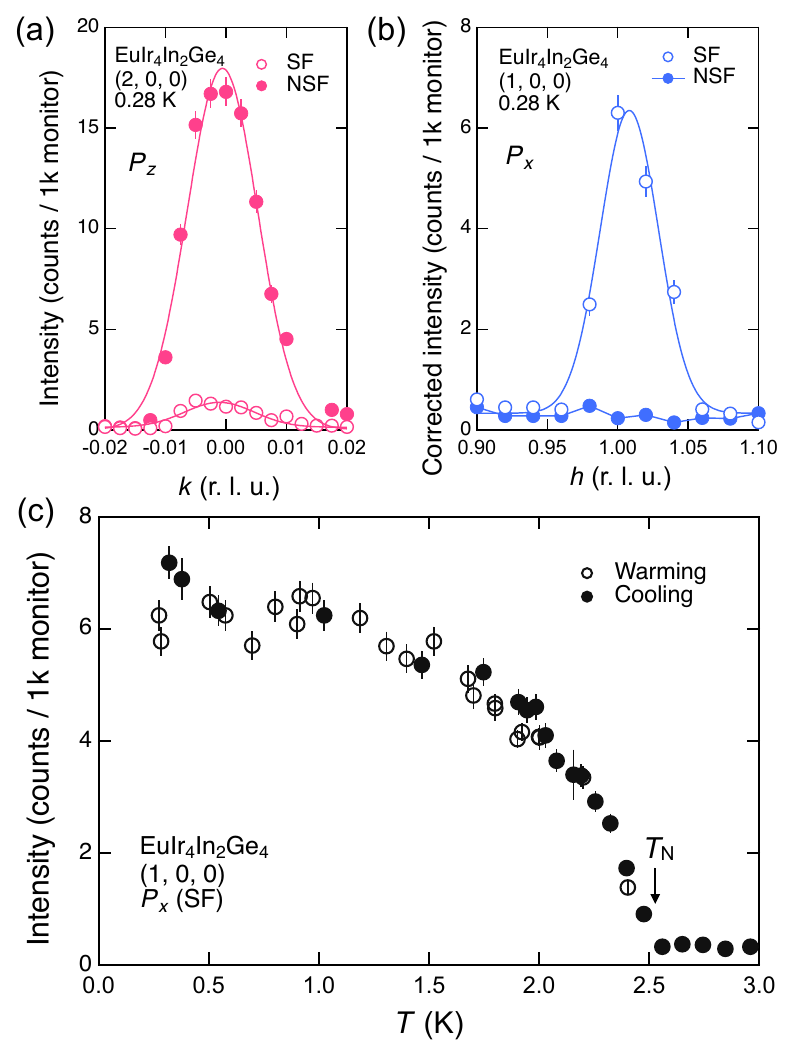}
    \caption{(Color online) Polarized neutron scattering profiles of (a) the 2~0~0 nuclear reflection and (b) the 1~0~0 magnetic reflection measured in the $P_x$ mode at 0.28~K. (c) Magnetic scattering intensity of the 1~0~0 reflection in the spin flip channel for $P_x$ as a function of temperature measured on warming and cooling processes.}
    \label{fig:Px}
\end{figure}

Figure \ref{fig:Px}(a) shows the peak profile of the 2~0~0 nuclear reflection of \eiig~measured in the $P_x$ mode.
A flipping ratio of $R = 20.6$ was obtained by comparing the integrated intensities of the SF and NSF components derived from Gaussian fitting.
Using this value, the intensity of the 1~0~0 Bragg reflection measured at 0.28 K was corrected, as shown in Fig. \ref{fig:Px}(b).
Only the SF signal exhibits a distinct peak, confirming the purely magnetic origin of this reflection, in accordance with Eqs. (1) and (2).

The temperature dependence of the corrected SF intensity is presented in Fig.~\ref{fig:Px}(c).
The intensity clearly develops below {\tn =} 2.5~K, indicating that the magnetic reflection is associated with the bulk phase transition \cite{Nakachi2024-yg}.
No detectable hysteresis was observed between warming and cooling processes, suggesting a second-order transition consistent with the ${\lambda}$-type anomaly observed in the specific-heat data \cite{Nakachi2024-yg}.

\begin{figure}[h]
\centering
        \includegraphics[width=8cm]{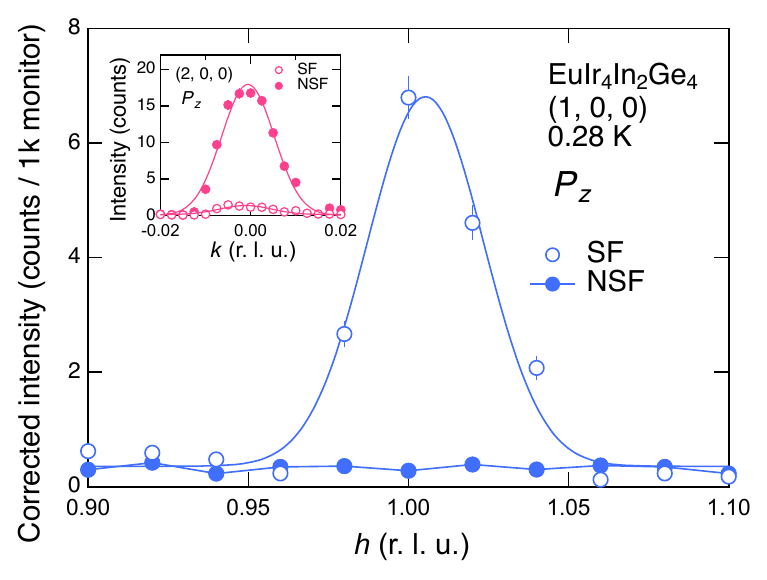}
    \caption{(Color online) Polarized neutron scattering profile of the 1~0~0 reflection measured in $P_z$ mode at 0.28~K. The intensity was corrected using the flipping ratio observed for the 2~0~0 nuclear Bragg reflection shown in the inset.}
    \label{fig:Pz}
\end{figure}

In the same manner, the 1~0~0 peak profile was investigated in the $P_z$ mode, as shown in Fig. \ref{fig:Pz}.
The flipping ratio used for $P_z$ correction was $R = 14.7$, determined from the 2~0~0 nuclear reflection displayed in the inset. 
The 1~0~0 reflection was again observed exclusively in the SF channel.
According to Eqs. (3) and (4), the absence of NSF scattering in the $P_z$ mode unambiguously indicates that the ordered magnetic moments have no out-of-plane (i.e., $c$-axis) component; therefore, the moments lie within the basal plane.

\begin{figure}[t]
\centering
        \includegraphics[width=6.0cm]{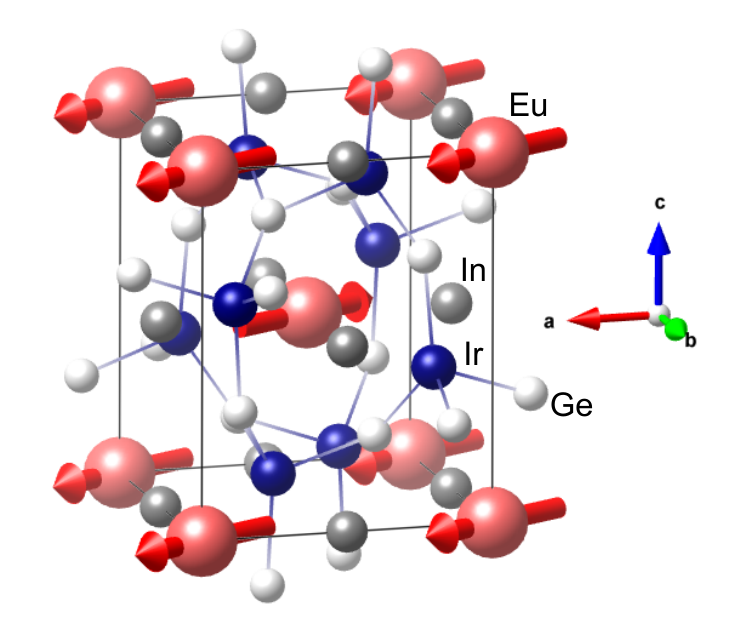}
    \caption{(Color online) Possible magnetic structure of {\eiig} in the antiferromagnetic ordered state below {\tn}. The in-plane direction of magnetic moments is not determined. The graphic was generated using the VESTA software \cite{Momma2011-eg}.}
    \label{fig:structure}
\end{figure}

Hereafter, the magnetic structure of {\eiig} below {\tn} is discussed.
The present experimental results establish a magnetic propagation vector of $\bm{q}=(1, 0, 0)$.
Together with the absence of spontaneous magnetization in bulk magnetization measurements \cite{Nakachi2024-yg}, this indicates that the Eu$^{2+}$ moments at the corner and body-center sites of the unit cell are aligned antiparallel to each other.
In addition, polarized neutron diffraction further demonstrates that the ordered moments lie within the basal plane.
The resulting magnetic structure model is illustrated in Fig. \ref{fig:structure}.

The in-plane orientation of the magnetic moments, however, cannot be uniquely determined because of magnetic domains arising from the equivalence of the tetragonal $a$ and $b$ axes.
Irreducible representation analysis, which is often a powerful tool for constraining magnetic structure models, unfortunately does not provide further insight in the present case.
Assuming a Landau-type phase transition, two types of ordered-moment components are allowed: $(0, 0, w)$ and $(u, v, 0)$, corresponding to irreducible representations $\Gamma_3$ and $\Gamma_5$, respectively \cite{Aroyo2011-kw}.
The former is immediately excluded by the experimental observation that the ordered moments have no $c$-axis component.
Consequently, the latter, which permits an arbitrary orientation within the $ab$ plane, is consistent with the present experimental results.
Complementary microscopic probes, such as NMR and  M\"ossbauer spectroscopies, may provide valuable information to determine the precise in-plane orientation of the ordered moments.

In contrast to a Dresselhaus-type spin splitting of the conduction electrons, the magnetic order formed by localized \ce{Eu^2+} $4f$ moments adopts a simple collinear structure with the moments lying within the $c$-plane.
This is consistent with the observed easy-plane anisotropy in the magnetization measurements \cite{Nakachi2024-yg}. 
The absence of canting indicates that antisymmetric exchange interactions play no significant role in the present system.

For comparison, in the tetragonal melilite \ce{Sr2MnSi2O7}, a canted spin structure is stabilized by the combination effects of superexchange and the Dzyaloshinskii-Moriya (DM) interaction arising from the Dresselhaus-type crystal symmetry \cite{Nambu2024-py}. 
In $f$-electron magnets, however, magnetic interactions are typically dominated by the Runderman-Kasuya-Kittel-Yoshida (RKKY) interaction mediated by conduction electrons, rather than by superexchange.
An supplementary DM interaction is considered to stabilize a helical magnetic order in the polar magnet EuNiGe$_3$ \cite{Matsumura2024-ic}.
At present, the fundamental magnetic exchange interactions in {\eiig} have not yet been clarified. Further investigation, including theoretical analysis, will be necessary to elucidate the microscopic mechanism governing its magnetic order.

Noncentrosymmetric magnets with Rashba-type antisymmetric spin-orbit coupling, such as $RTX_3$ ($T$: transition metals, $X$: Si, Ge) and \ce{CePt3Si}, exhibit pronounced band splitting at the Fermi surface \cite{Hashimoto2004-pt, Kawai2008-rg, Onuki2012-om}.
Nevertheless, collinear magnetic structures have also been reported in several of these systems \cite{Metoki2005-mx,Fak2017-hn, Smidman2013-an}.
To date, no systematic correspondence between the form of band splitting and the resulting magnetic structure has been established.

In conclusion, the magnetic order of the noncentrosymmetric tetragonal antiferromagnet \eiig~was elucidated by neutron diffraction experiments.
A wide $Q$ range surveyed using a time-of-flight diffractometer revealed magnetic reflections characterized by the propagation vector $\bm{q}=(1, 0, 0)$, which breaks the body-centered symmetry.
Subsequent polarized neutron scattering measurements on a triple-axis spectrometer demonstrated that the ordered magnetic moments lie within the basal plane.
These results establish a magnetic structure in which the in-plane moments at the corner and body-center sites are antiparallel.
This structure does not reflect the broken inversion symmetry of the crystal lattice, implying that the localized $4f$ moments are effectively separated from the conduction electrons exhibiting Dresselhaus-type spin splitting.

\begin{acknowledgments}
We acknowledge valuable discussions with A. Miyake, D. Aoki, and M. Kimata. 
This work was supported by JSPS KAKENHI grants Nos. JP23H04867, JP23H04870, JP22K03517, and JP21H04987. 
The neutron experiments were performed at MLF of J-PARC and JRR-3 under Proposal No. 2023I0018 and D1097, respectively.
\end{acknowledgments}

\vspace{1em}
\noindent\textsuperscript{$\dagger$}\texttt{tabata.chihiro@jaea.go.jp}

\bibliography{EIIG_reference.bib}
\end{document}